\documentclass[sigconf,natbib=true,anonymous=false]{acmart}

\AtBeginDocument{%
  \providecommand\BibTeX{{%
    \normalfont B\kern-0.5em{\scshape i\kern-0.25em b}\kern-0.8em\TeX}}}

\copyrightyear{2022} 
\acmYear{2022} 
\setcopyright{acmlicensed}\acmConference[SIGIR '22]{Proceedings of the 45th International ACM SIGIR Conference on Research and Development in Information Retrieval}{July 11--15, 2022}{Madrid, Spain}
\acmBooktitle{Proceedings of the 45th International ACM SIGIR Conference on Research and Development in Information Retrieval (SIGIR '22), July 11--15, 2022, Madrid, Spain}
\acmPrice{15.00}
\acmDOI{10.1145/3477495.3531769}
\acmISBN{978-1-4503-8732-3/22/07}

\usepackage{booktabs}
\usepackage{graphicx}
\usepackage{enumitem}

\usepackage{lipsum}
\usepackage{float}
\usepackage{subfigure}
\usepackage{multirow,makecell}
\usepackage{pifont}
\usepackage{caption}

\begin{document}
\fancyhead{}
\title{Zero-shot Dense Retrieval for Conversational Search}
\title{Zero-shot Query Contextualization for Conversational Search}

\author{Antonios Minas Krasakis}
\email{a.m.krasakis@uva.nl}
\affiliation{%
  \institution{University of Amsterdam}
  \country{The Netherlands}
}
\author{Andrew Yates}
\email{a.c.yates@uva.nl}
\affiliation{%
  \institution{University of Amsterdam}
  \country{The Netherlands}
}
\author{Evangelos Kanoulas}
\email{e.kanoulas@uva.nl}
\affiliation{%
  \institution{University of Amsterdam}
  \country{The Netherlands}
}

\renewcommand{\shortauthors}{Krasakis, et al.}

\begin{abstract}
Current conversational passage retrieval systems cast conversational search into ad-hoc search by using an intermediate query resolution step that places the user's question in context of the conversation. While the proposed methods have proven effective, they still assume the availability of large-scale question resolution and conversational search datasets.
To waive the dependency on the availability of such data, we adapt a pre-trained token-level dense retriever on ad-hoc search data to perform conversational search with no additional fine-tuning. The proposed method allows to contextualize the user question within the conversation history, but restrict the matching only between question and potential answer.
Our experiments demonstrate the effectiveness of the proposed approach. We also perform an analysis that provides insights of how contextualization works in the latent space, in essence introducing a bias towards salient terms from the conversation.
\end{abstract}




\maketitle

\section{Introduction}

The introduction of commercial voice assistants along with advances in natural language understanding have enabled users to interact with retrieval systems in richer and more natural ways through conversations. While those interactions can ultimately lead to increased user satisfaction, they are inherently complex to handle as they require an understanding of the entire dialogue semantics by the retrieval system. Hence, the retrieval of relevant passages within the context of a conversation has risen as a promising research direction ~\cite{dalton2020trec,krasakis2020analysing,borisov2021keyword, meng2021initiative}.

Understanding the semantics of language has been empowered by the availability of large-scale datasets in a variety of tasks~\cite{choi2018quac,nguyen2016ms,yates2021pretrained}, which are lacking when it comes to conversational retrieval.
Constructing a large and diverse conversational retrieval dataset can be quite challenging.
Conversational queries are tail queries. As conversations evolve, multi-turn queries are likely to be unique and therefore cannot be aggregated for anonymisation, making it unlikely that publicly available resources can be built from real user interactions.
Therefore, datasets need to be built using human experts in controlled environments. 
However, this approach leads to small-scale datasets~\cite{dalton2020trec,dalton2020trec1,dalton2022trec,10.1145/3432726} and requires explicit conversation development instructions which bias the nature of the constructed dataset and hurt the generalizability of models to new types of conversations~\cite{adlakha2021topiocqa}. 

On the other hand there is a plethora of data resources for ad-hoc retrieval, e.g.\ ~\citet{craswell2020orcas}. Therefore, most conversational retrieval approaches so far introduce a query rewriting step, which essentially decomposes the conversational search problem into a query resolution problem and an ad-hoc retrieval problem. 
Regarding query resolution, the majority of methods perform an explicit query re-write attempting to place the user's question in the context of the conversation, by either expanding queries with terms from recent history ~\cite{voskarides2020query}, or rewriting the full question using a sequence-to-sequence model ~\cite{vakulenko2021comparison,lintrec,ferreira2021open,lin2021multi,yu2020few}. 
~\citet{yu2021few} learns to better encode the user's question in a latent space so that the learnt embeddings are close to human rewritten questions, while ~\citet{lin2021contextualized} uses human rewritten questions to generate large-scale pseudo-relevance labels and bring the user's question embeddings closer to the pseudo-relevant passage embeddings.
In all cases, supervision is necessary and it is performed against CANARD~\cite{elgohary2019can}, which consists of $40K$ synthetic resolutions of conversational questions.
The only approaches that do not use supervision simply expand the user's question by extracting general informative terms from the conversation history~ \cite{lin2021multi,borisov2021keyword}.

\begin{figure}
\centering
\includegraphics[scale=0.4]{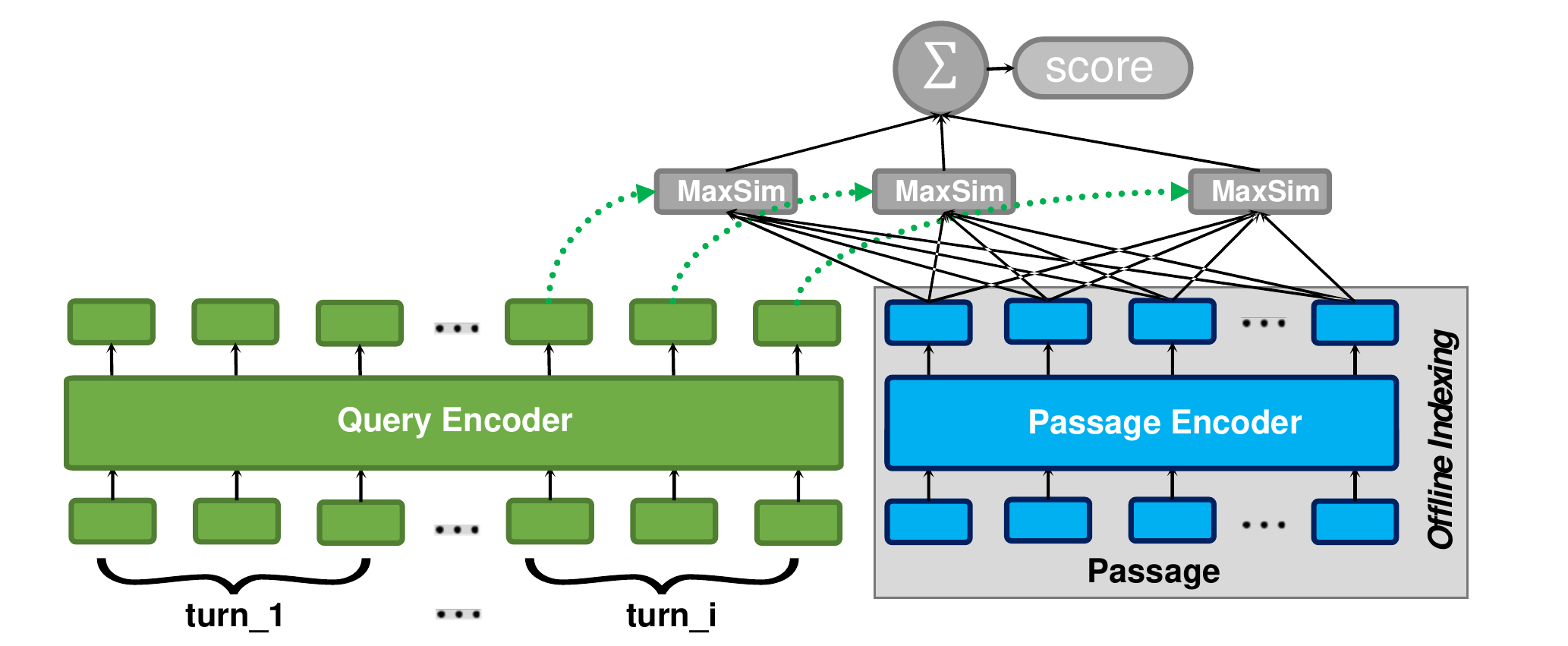}
\caption{Zero-shot Conversational Dense Retriever (figure adapted from ~\cite{khattab2020colbert} with permission)}\label{fig:0sConvDR}
\vspace*{-0.3cm}
\end{figure}

In this work we pose the following key research question: \emph{to what extent can we transfer knowledge from ad-hoc retrieval to the domain of conversational retrieval, where data scarcity is and will likely remain an imminent problem?}
To answer this question we adapt ColBERT, the state-of-the-art BERT-based token-level dense retriever pre-trained on ad-hoc search data. We propose \textit{\textbf{Ze}ro-shot \textbf{Co}nversational \textbf{Co}ntextualization} ($ZeCo^2$), a variant of ColBERT which on one hand contextualizes all embeddings within the conversation, but on the other hand matches only the contextualized terms of the last user's question with potential answers (Figure \ref{fig:0sConvDR}). As such our approach is zero-shot in the conversational domain, that is, it does not use any conversational search data, neither rewritten queries nor relevance judgements, to retrieve relevant passages. 
It is also different from the aforementioned unsupervised keyword extraction works, since it focuses on contextualizing embeddings rather than adding terms to queries.
In this work we aim to answer the following research questions:
\begin{enumerate}[label=\textbf{RQ\arabic*},leftmargin=*]
	\item Can zero-shot contextualization of conversational questions improve dense passage retrieval?
	\item How does zero-shot contextualization change the last turn's question embeddings?
	\item How is zero-shot contextualization affected by the anaphora phenomena found in conversations?
\end{enumerate}

%
To the best of our knowledge this is one of the few efforts for zero-shot conversational search. Our approach remains agnostic and unbiased to small conversational datasets and it can prove particularly useful in privacy-sensitive settings (eg. medical domain), where annotating rewrites of conversational questions is not an option. %
Further, our method is orthogonal to and can be applied in combination with existing query resolution techniques.

We open-source our code for reproducibility and future research purposes \footnote{https://github.com/littlewine/ZeCo2}.

\section{Methodology}

In this section, we describe our zero-shot dense retriever for conversational retrieval. Our approach consists of two main components: an encoder that produces token embeddings of a document or query and a matching component that compares query and document token embeddings to produce a relevance score.

\subsection{Task \& Notation}

Let $q_t$ be the user utterance/query to the system at the $t$-th turn, and $p_t$ the corresponding canonical passage response provided by the dataset.
We formulate our passage retrieval task as follows: Given the last user utterance $q_t$ and the previous context of the conversation at turn $t$: $ctx_t = (q_0, p_0, ... , q_{t-1}, p_{t-1})$, we produce a ranking of k passages $R_{q_t} = (p_t^1, p_t^2, ... , p_t^k)$ 
from a collection $C$ that are most likely to satisfy the users' information need.

\subsection{Token-level Dense Retrieval}
In this section we briefly describe ColBERT~\cite{khattab2020colbert}, a dense retrieval model that serves as our query and document encoder $f_{Enc}$. In contrast to other dense retrievers that construct global query and document representations (eg. DPR~\cite{karpukhin2020dense} or ANCE~\cite{xiong2020approximate}), ColBERT generates embeddings of all input tokens. This allows us to perform matching at the token-level.
%
To generate token embeddings, ColBERT passes each token through multiple attention layers in a transformer encoder architecture, which contextualizes each token with respect to its surroundings ~\cite{vaswani2017attention,devlin2018bert}. 
We use $E_q$ to denote the embeddings produced for a query $q$ and $E_d$ to denote the embeddings produced for a document $d$.
To compute a query-document score, ColBERT performs a soft-match between the embeddings of a query token $w_q$ and a document token $w_d$ by taking their inner product.
Specifically, each query token is matched with the most similar document token and the summation is computed:

\begin{equation}\label{colbert-matching}
Score(q, d) := \sum_{w_q \in q} \max_{w_d \in d} E_{w_q} \cdot E^T_{w_d}
\end{equation}

\subsection{Conversational Token-level Dense Retrieval}\label{method-ConvTokDR}

Our approach extends the idea of token contextualization to multi-turn conversations. When dealing with conversations, it is crucial for each turn to be contextualized with respect to the conversation, because conversational queries have continuity and often contain phenomena of ellipsis or anaphoras to previous turns ~\cite{vakulenko2021question,yu2020few,voskarides2020query}.

In practice, ColBERT serves as our query and document encoder $f_{Enc}$. We encode documents in the usual way. 
However, to encode a query at turn $t$, we concatenate the conversational context $ctx_t$ with the last query utterance $q_t$ before generating contextualized query token embeddings $E^*_{q^t}$.

\begin{equation}\label{eqn:embedding-contextualization}
    E^*_{q^t} := f_{Enc}(ctx_t \circ [SEP] \circ q_t )
\end{equation}

While $E^*_{q^t}$ constitutes token embeddings of the entire conversation (i.e., the input to $f_{Enc}$), our goal is to perform ranking based on only tokens from the last utterance $q_t$.
To do so, we (1) replace $E_{w_q}$ with $E^*_{w_q}$ in the token-level matching function (Eq \ref{colbert-matching}) and (2) compute the score as $Score(q_t, d)$, so that only query tokens from the last turn contribute to it.

\begin{equation}
Score(q_t, d) := \sum_{w_q \in q_t} \max_{w_d \in d}{ E^*_{w_q} \cdot E^T_{w_d}}
\end{equation}

Note that, this approach of contextualizing $q_t$ with respect to the conversation history $ctx_t$ avoids the need for resolution supervision from conversational tasks. 
Instead, it relies on the pre-training of three different tasks: (a) Masked Language Modelling, (b) next sentence prediction tasks (pre-training of BERT ~\cite{devlin2018bert}) and the (c) ad-hoc ranking task (pre-training of ColBERT~\cite{khattab2020colbert}).



\begin{table*}[th!]
\begin{tabular}{llc||ll|ll|ll}
\multirowcell{2}{base-\\retriever} &  \multirow{2}{*}{variant}   & \multirowcell{2}{zero-\\shot}  & \multicolumn{2}{c}{CAsT'19}     & \multicolumn{2}{c}{CAsT'20}     & \multicolumn{2}{c}{CAsT'21}     \\
 &  &  & NDCG@3         & R@100         & NDCG@3         & R@100         & NDCG@3         & R@100         \\
\hline\hline

\multirow{4}{*}{ColBERT} & last-turn $^a$ & 
\ding{52} & 0.214          & 0.157     & 0.155          & 0.124      & 0.140     & 0.154    \\
& all-history $^b$ & 
\ding{52} & 0.190    & 0.165    & 0.150  & 0.166      & \textbf{0.237} & 0.265          \\
& \begin{tabular}[c]{@{}l@{}}ZeCo$^2$ \textit{(ours)}\end{tabular} & 
\ding{52} & 0.238 $^{b}$ & \textbf{0.216} $^{a,b,c}$ & \textbf{0.176} $^{b}$ & \textbf{0.200} $^{a,b,c}$ & 0.234 $^{a}$& \textbf{0.267}  $^{a}$ \\
& human  &  & 0.430           & 0.363          & 0.443          & 0.408           & 0.431          & 0.403         \\
\hline
\multirow{3}{*}{ConvDR~\cite{yu2021few}} & zero-shot $^c$ & 
\ding{52} & \textbf{0.247}          & 0.183     & 0.150      & 0.150      & --           & --       \\
& few-shot  & & 0.466          & 0.362      & 0.340        & 0.345       & 0.361           & 0.376       \\
& human   & & 0.461     & 0.389 & 0.422   & 0.454    & 0.548  & 0.451      \\
\hline

\end{tabular}
\caption{Effectiveness of zero-shot embedding contextualization on TREC-CAsT datasets.
Bold font indicates the best zero-shot performing model. Superscripts indicate statistically significant improvements (paired t-test, $p-value<0.05$) of $ZeCo^2$ over zero-shot models: \textit{last-turn $^a$, all-history $^b$} and \textit{ConvDR zero-shot $^c$} }
\vspace*{-0.6cm}
\label{tab:results-contextualization}
\end{table*}

\section{Experimental Setup}
\subsection{Datasets and Evaluation}
We test our approach on the TREC CAsT '19, '20 and '21~\cite{dalton2020trec,dalton2020trec1,dalton2022trec} datasets. Each dataset consists of about 25 conversations, with an average of 10 turns per conversation. CAsT '20 and '21 include canonical passage responses to previous questions, that the user can refer to or give feedback.
The corpus consists of the MSMarco Passages and Documents, Wikipedia and Washington Post news articles~\cite{nguyen2016ms,petroni2020kilt, dietz2017trec}. TREC CAsT '19 and '20 includes relevance judgements at passage level, whereas CAsT '21 at the document level. For CAsT '21, we split the documents into passages and score each document based on its highest scored passage ($MaxP$ \cite{dai2019deeper}).

To quantify retrieval performance we use two metrics: NDCG@3 and Recall, R@100. The former quantifies effectiveness at the top ranks, which is important for a user, while the latter expresses the ability of a first-stage ranker to retrieve relevant passages that can be later re-ranked with a more effective second-stage ranker.

\subsection{Methods \& Baselines}\label{sec:methods-baselines}
\subsubsection*{ColBERT} ColBERT was trained to contextualize tokens through (a) the self-supervision of BERT's masked language model and next-sentence prediction~\cite{devlin2018bert} and (b) the training to optimize ad-hoc ranking~\cite{khattab2020colbert}. In our experiments, we use the weights of a ColBERT retriever pre-trained on the MSMarco passage ranking dataset~\cite{nguyen2016ms}. 
We use ColBERT v1, while our method remains applicable to v2~\cite{santhanam2021colbertv2}, where the main novelty is optimizations to reduce the index size.
To avoid any spill-over effects we perform matching only on the query tokens; we deactivate matching on the $CLS$, query indicator ($[Q]$) and expansion tokens used in the original work~\cite{khattab2020colbert}.

\subsubsection*{Baselines}
To assess the effect of our contextualization method $ZeCo^2$, we compare with the following baselines: \\ \vspace*{-0.3cm}\\
$ColBERT$-based:
\begin{itemize}
    \item \textit{last-turn}: embeddings are not contextualized in the conversation (i.e., $ctx_t = \emptyset$ in Eq. (\ref{eqn:embedding-contextualization})).
    
    \item \textit{all-history}: embeddings are contextualized in conversation, and the matching function includes terms across the entire history (i.e., $Score(ctx_t \circ q_t, d)$).
    \item \textit{human}: oracle, using human rewrites
\end{itemize}
$ConvDR$-based ~\cite{yu2021few}:
\begin{itemize}
    \item \textit{zero-shot}: no supervision from conversational tasks/data
    \item \textit{few-shot}: trained with Knowledge-Distillation on query rewrites 
    \item \textit{human}: oracle, using human rewrites
\end{itemize}

\section{Results and Analysis}

To answer \textbf{RQ1}, which asks whether the last user utterance (question) can be effectively contextualized with respect to the conversation history, we compare the performance of the non-contextualized utterance (\textit{last-turn}) with our contextualized approach ($ZeCo^2$) in Table \ref{tab:results-contextualization}. It is clear that contextualization helps in all cases, especially in terms of \textit{Recall} with relative improvements of $37\%$ - $73\%$.
We further observe that our approach significantly outperforms the \textit{all-history} baseline, which uses the entire conversation as the query, in the first two datasets and yields comparable performance on CAsT'21. We hypothesize that the baseline's improved performance on CAsT'21 is due to its document-level annotations, with one document satisfying multiple turns of the conversation.
We also observe that \textit{all-history} performs worse than \textit{last-turn} regarding $NDCG@3$ but better regarding $R@100$.
Furthermore, $ZeCo^2$ outperforms the zero-shot ConvDR in most cases, especially with respect to $Recall$. Last, while the supervised versions of ConvDR clearly outperfom $ZeCo^2$ in $NDCG@3$, $ZeCo^2$ remains competitive in terms of $Recall$. 

Next we consider the effect of contextualization of the user's query by looking into how this changes the last turn's query embeddings so that we answer \textbf{RQ2}.

\textit{What are the most influenced terms?} To assess the effect of conversational contextualization ($ZeCo^2$), we measure the token embedding changes in the user's query and report the terms with the largest average change in Table~\ref{table:mostchangingterms}. We define token embedding change as the cosine distance between a term before and after contextualization:
$\Delta \vec{tok} = 1 - cos(\vec{tok}_{ZeCo^2} , \vec{tok}_{last-turn})$. 
We observe that terms indicating anaphora ('they', 'it', etc.), punctuation symbols and special tokens are the ones most influenced. This is expected, since users often use anaphoras referring to previous conversation rounds. Regarding punctuation and special tokens, one plausible explanation is that a global representation of a turn is aggregated in those tokens after contextualization.
\begin{table}[H]

\begin{tabular}{lrr}
\toprule
{} &  frequency &  avg($\Delta \vec{tok}$) \\
token  &              &             \\
\midrule
they  &           60 &       0.501 \\
it    &          196 &       0.480 \\
$[SEP]$ &          934 &       0.464 \\
?     &          873 &       0.458 \\
that  &           52 &       0.440 \\
.     &          192 &       0.424 \\
...   &     ...    &  ... \\\hline
macro-avg   &           -- &       0.185 \\
micro-avg   &           -- &       0.323 \\

\bottomrule
\end{tabular}

\caption{Average change of frequent token embeddings after zero-shot contextualization (all CAsT datasets).}
\label{table:mostchangingterms}
\end{table}



\begin{table*}[t]

\begin{tabular}{lll|r|lc|lc}
\multirow{3}{*}{Utterance}                                                                          & \multirow{3}{*}{\begin{tabular}[c]{@{}l@{}}Human \\ resolution\end{tabular}} & \multirow{3}{*}{\begin{tabular}[c]{@{}l@{}}$\Delta$ Anaphora\\ embedding\end{tabular}} & \multirow{3}{*}{$\Delta$ Recall} 
&
\multicolumn{2}{c}{\begin{tabular}[c]{@{}c@{}}closest match\\ (non-contextualized)\end{tabular}}
& 

\multicolumn{2}{c}{\begin{tabular}[c]{@{}c@{}}closest match\\ (contextualized)\end{tabular}}\\
&                                                                              &                                                                                                     &                       
& term & similarity
& term & similarity \\\hline\hline

\begin{tabular}[c]{@{}l@{}}what is the \\ first sign of \underline{\textbf{it}} ?\end{tabular}                         & \begin{tabular}[c]{@{}l@{}}throat\\ cancer\end{tabular}                      & 0.52                                                                                                & +0.31                 & \begin{tabular}[c]{@{}l@{}}tell me about \\ lung \underline{\textbf{cancer}.}\end{tabular}     & 0.48       & \begin{tabular}[c]{@{}l@{}}what causes \\ throat \underline{\textbf{cancer}} ?\end{tabular}                 & 1          \\\hline
\begin{tabular}[c]{@{}l@{}}What is the role of \\ positivism in \underline{\textbf{it}}?\end{tabular}                  & sociology                                                                    & 0.44                                                                                                & +0.67                 & \begin{tabular}[c]{@{}l@{}}what is taught in \\ \underline{\textbf{sociology}} ?\end{tabular}                         & 0.55       & \begin{tabular}[c]{@{}l@{}}what is taught in \\ \underline{\textbf{sociology}} ?\end{tabular}               & 1          \\\hline

\begin{tabular}[c]{@{}l@{}}what technological \\ developments enabled \underline{\textbf{it}} ?\end{tabular}           & \begin{tabular}[c]{@{}l@{}}popular\\ music\end{tabular}                      & 0.54   & +0.42                 & \begin{tabular}[c]{@{}l@{}}... the origins of\\ popular \underline{\textbf{music}} ?\end{tabular}                     & 0.46       & \begin{tabular}[c]{@{}l@{}}... the origins of\\ popular \underline{\textbf{music}} ?\end{tabular}           & 1          \\\hline

\begin{tabular}[c]{@{}l@{}}what is the \\ evidence for \underline{\textbf{it}} ?\end{tabular}                          & \begin{tabular}[c]{@{}l@{}}bronze age\\ collapse\end{tabular}                & 0.36       & 0.00                 & \begin{tabular}[c]{@{}l@{}}tell me about the \\ bronze \underline{\textbf{age}} collapse.\end{tabular}                & 0.64       & \begin{tabular}[c]{@{}l@{}}tell me about the bronze \\ age \underline{\textbf{collapse}}.\end{tabular}    & 1          \\\hline

\begin{tabular}[c]{@{}l@{}}why did ben franklin want \underline{\textbf{it}} \\ to be the national symbol?\end{tabular} & turkeys                                                                      & 0.22                                                                                                & +0.29                 & \begin{tabular}[c]{@{}l@{}}where are \\ turkeys from \underline{\textbf{?}}\end{tabular}                              & 0.55       & \begin{tabular}[c]{@{}l@{}}where are \\ turkeys from \underline{\textbf{?}}\end{tabular}                    & 0.85       \\\hline

\begin{tabular}[c]{@{}l@{}}what is \underline{\textbf{it}} \\  about?\end{tabular} & \begin{tabular}[c]{@{}l@{}}neverending \\  story film\end{tabular}  & 0.39                                                                                                & +0.12                 & \begin{tabular}[c]{@{}l@{}}the neverending \\ story film \underline{.}\end{tabular}                               & 0.58       & \begin{tabular}[c]{@{}l@{}}the neverending \\ story film \underline{\textbf{.}}\end{tabular}                    & 0.88       \\\hline\hline


\end{tabular}

\caption{Examples of best term matches of anaphora terms in conversation history (before/after query contextualization).}
\label{tab:closest-terms-conversation}
\end{table*}

\vspace{-0,5cm}
\textit{How do terms change when contextualized?} To illustrate how contextualization changes term embeddings, in Table \ref{tab:closest-terms-conversation} we focus on a highly influenced anaphora term ('it') and match it to the most similar token embeddings from the conversation history.
We observe that in certain cases, zero-shot contextualization resolves anaphoras successfully, bringing anaphora embeddings very close to the referred term (``sociology'', ``popular music'', ``throat cancer'').
%
%
The first row shows one noteworthy example where the matching term is always \textit{cancer}, but contextualization allows it to resolve to the correct embedding of \textit{throat cancer} instead of \textit{lung cancer}.
%
Lastly, we see cases where embeddings come closer to punctuation symbols, indicating that those might preserve some sort of global query representation, or a multi-token concept (e.g., ``the neverending story film .'').

\begin{figure}
\centering
\vspace*{-0.5cm}
\includegraphics[scale=0.55]{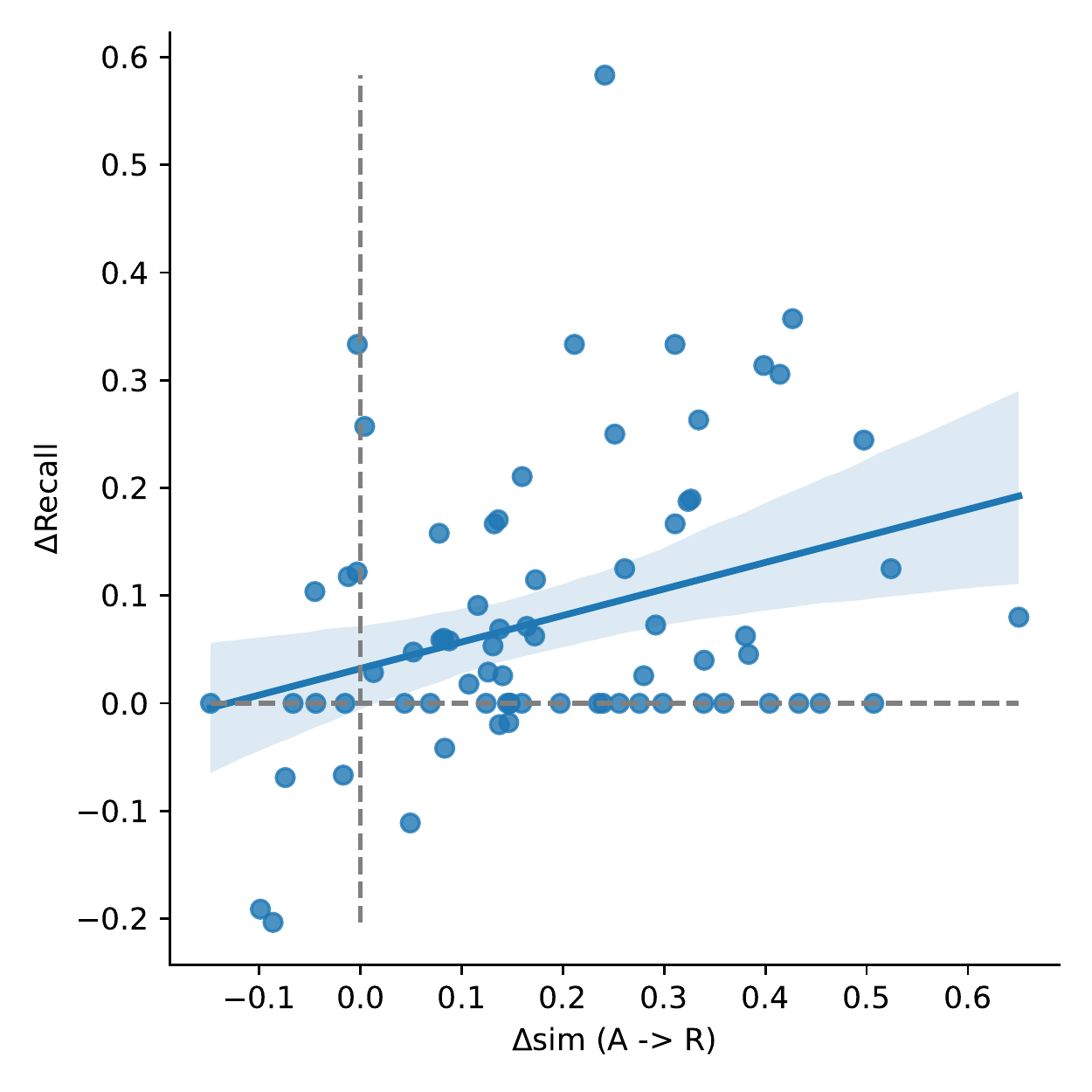}
\vspace*{-0.4cm}
\caption{Correlation between $\Delta Recall$ and similarity change of anaphoras towards resolutions (CAsT '19)}\label{fig:correlation-Drecall}
\vspace*{-0.3cm}
\label{fig:histograms}
\end{figure}

To answer \textbf{RQ3}, we quantitatively explore whether contextualization brings anaphora terms closer to their corresponding resolutions and how this affects ranking. 
Bringing anaphora terms closer to resolutions is crucial for conversational search.
We identify those terms automatically using the human rewrites 
and define the effect of contextualization in bringing anaphora embeddings ($\vec{A}$) closer to resolution embeddings ($\vec{R}$) as: \vspace{-0.4cm}

\[\Delta sim (\vec{A} \rightarrow \vec{R}) = sim(\vec{A}_{ZeCo^2}, \vec{R} )
- sim(\vec{A}_{last-turn}, \vec{R} )\] 
where anaphoras are contextualized within the last turn ($\vec{A}_{last-turn}$) or the entire conversation ($\vec{A}_{ZeCo^2}$). We encode resolutions ($\vec{R}$) independently to ensure they retain their original representations. On queries with multi-token anaphoras or resolutions, we pick the highest match.
In Figure \ref{fig:correlation-Drecall} we observe the scatter plot of this $\Delta sim$ against $\Delta Recall$.
In most cases, contextualization improves $Recall$ ($\Delta Recall>0$) and brings anaphoras closer to resolutions ($\Delta sim>0$). Further, Recall correlates with this change in similarity towards resolutions (Pearson's $R = 0.31$, $p-value=0.005$). 

\begin{table}[]
\vspace{-0.2cm}
\begin{tabular}{l||ll}
   & $\vec{R}$ (resolution) & $\vec{Rnd}$ (random term)\\\hline\hline
  $\vec{A}_{last-turn}$ & 0.178           & 0.255       \\
 $\vec{A}_{ZeCo^2} $   & 0.372           & 0.204      
\end{tabular}
\vspace{0.1cm}
\caption{Embedding similarities between anaphoras and resolution or random terms from the conversations (CAsT '19).}
\vspace*{-0.7cm}
\label{tab:anaphora-resolution-random-similarities}
\end{table}


To examine whether anaphora terms coming closer to resolutions is simply a by-product of being encoded together, we measure similarities between anaphoras ($\vec{A}$) and (a) resolution terms ($\vec{R}$) vs (b) random terms from the same conversation ($\vec{Rnd}$) in Table \ref{tab:anaphora-resolution-random-similarities}. 
For consistency, we also encode random terms independently.
When anaphoras are contextualized within only the \textit{last-turn}, they are more similar to random terms than to resolutions on average.
However, our method ($ZeCo^2$) brings anaphoras closer to their resolutions, while pushing them away from other (random) words from the same conversation. This confirms that resolutions have a high impact on contextualizing anaphoras, in contrast to other random conversation words. 
The mechanism behind this effect requires further investigation. It could be that simply the lower frequency of resolution terms has an effect here, but it is also possible that pre-trained transformers have certain co-reference resolution capabilities (eg. by relating 'it' to a noun).
Regardless, it is evident that our method induces a bias towards salient terms from the conversation, leading to improved ranking performance.

\section{Conclusions}
In this paper, we explore the possibility of performing conversational search in a zero-shot setting, by contextualizing the last user query with respect to the conversation history. We show that this method is highly effective for first-stage ranking, yielding consistent and significant improvements in $R@100$.
Further, it is suitable for privacy-sensitive settings, and can be combined with existing query rewriting techniques.
In addition, we shed light into how zero-shot contextualization changes the last turn embeddings and show that biasing them towards the previous conversation can help retrieval, since it brings them closer to the conversation topic and salient terms.
For future work we aim to explore zero-shot re-ranking and extend this work to few-shot training.

\begin{acks}
This research was supported by
the NWO Innovational Research Incentives Scheme Vidi (016.Vidi.189.039),
the NWO Smart Culture - Big Data / Digital Humanities (314-99-301), and
the H2020-EU.3.4. - SOCIETAL CHALLENGES - Smart, Green And Integrated Transport (814961).
All content represents the opinion of the authors, which is not necessarily shared or endorsed by their respective employers and/or sponsors.
\end{acks}

\bibliographystyle{ACM-Reference-Format}
\bibliography{main}

\end{document}